# A multi-material level set-based topology optimization of flexoelectric composites

## Hamid Ghasemi[2], Harold S. Park[3], Timon Rabczuk[1, 2]

[1] Duy Tan University, Institute of Research & Development, 3 Quang Trung, Danang, Viet Nam
[2] Institute of Structural Mechanics, Bauhaus- Universität Weimar, Marienstraße 15, 99423 Weimar, Germany
[3] Dep. of Mechanical Engineering, Boston University, Boston, MA 02215, USA

**Abstract**

We present a computational design methodology for topology optimization of multi-material-based flexoelectric composites. The methodology extends our recently proposed design methodology for a single flexoelectric material. We adopt the multi-phase vector level set (LS) model which easily copes with various numbers of phases, efficiently satisfies multiple constraints and intrinsically avoids overlap or vacuum among different phases. We extend the point wise density mapping technique for multi-material design and use the B-spline elements to discretize the partial differential equations (PDEs) of flexoelectricity. The dependence of the objective function on the design variables is incorporated using the adjoint technique. The obtained design sensitivities are used in the Hamilton–Jacobi (H-J) equation to update the LS function. We provide numerical examples for two, three and four phase flexoelectric composites to demonstrate the flexibility of the model as well as the significant enhancement in electromechanical coupling coefficient that can be obtained using multi-material topology optimization for flexoelectric composites.

## 1. Introduction

In dielectric crystals with non-centrosymmetric crystal structure such as quartz and ZnO, electrical polarization is generated upon the application of uniform mechanical strain. This property of certain materials, which is known as piezoelectricity, is caused by relative displacements between the centers of oppositely charged ions. Details about the governing equations of piezoelectricity are available in [1-3].

---

Corresponding Authors: E-Mail: hamid.ghasemi@uni-weimar.de; parkhs@bu.edu; timon.rabczuk@uni-weimar.de

When the mechanical strain is applied non-uniformly, the inversion symmetry of a dielectric unit cell can be broken locally. Thus all dielectric materials, including those with centrosymmetric crystal structures, can produce an electrical polarization. This phenomenon is known as the flexoelectric effect, where the gradient of mechanical strain can induce electrical polarization in a dielectric solid. Readers are referred to [4, 5] and references therein for more details.

Micro-Nano electromechanical sensors and actuators made from piezoelectric or flexoelectric materials are increasingly used in applications such as implanted biomedical systems [6], environmental monitoring [7] and structural health monitoring [8]. These sensors and actuators are structurally simpler, provide high power density, and allow a broader range of material choice; however, their efficiency is usually low [9].

Conventional flexoelectric ceramics or single crystals are usually brittle and therefore susceptible to fracture. In contrast, flexoelectric polymers are flexible but exhibit weaker flexoelectric performance. Moreover, in a single flexoelectric structure, zones with high strain gradients contribute more to electrical energy generation. Thus, the efficiency of a sensor or an actuator fabricated entirely from a single flexoelectric material might be suboptimal. More interestingly, there exist significant opportunities to design piezoelectric composites without using piezoelectric constitutive materials while reaching piezoelectric performance that rivals that seen in highly piezoelectric materials [4]. Therefore, there are significant opportunities in being able to design multi-phase flexoelectric composites to bridge the gap between high flexoelectric performance and poor structural properties.

Topology optimization is a powerful approach that determines the best material distribution within the design domain. The present authors have already presented a computational framework for topology optimization of single material flexoelectric micro and nanostructures to enhance their energy conversion efficiency [10, 11]. The present research however, exploits the capabilities of topology optimization for the systematic design of a multi-phase micro and nano sensors and actuators made from different active and passive materials.

Contributions on piezoelectric structure design are often restricted by the optimal design of the host structure with fixed piezoelectric elements [12] or optimal design of piezoelectric elements with the given structure [13, 14]. Studies on multi-material design of piezoelectric structures are relatively rare. In fact, available works on multi-material topology optimization mostly employ Isotropic Material with Penalization (SIMP) technique [15]. Furthermore, we are not aware of any previous work studying the optimization of multi-material flexoelectric composites. By use of the level set method, this work provides a new perspective on simultaneous topology optimization of the elastic, flexoelectric and void phases within the design domain such that multi-material flexoelectric composites can be designed.

The remainder of this paper is organized thus: Section 2 summarizes the discretized governing equations of flexoelectricity, Section 3 contains the topology optimization based on the LSM, Section 4 provides numerical examples, and Section 5 offers concluding remarks.

## 2. A summary of the governing equations and discretization

A summary of the governing equations of the flexoelectricity is presented in this section. More details are available in [10, 16-18] and references therein. Accounting for the flexoelectricity, the enthalpy density, $\mathcal{H}$, can be written as

$$\mathcal{H}(\varepsilon_{ij}, E_i, \varepsilon_{jk,l}) = \frac{1}{2} C_{ijkl} \varepsilon_{ij} \varepsilon_{kl} - e_{ikl} E_i \varepsilon_{kl} - \mu_{ijkl} E_i \varepsilon_{jk,l} - \frac{1}{2} \kappa_{ij} E_i E_j \tag{1}$$

where $C_{ijkl}$ is the fourth-order elasticity tensor, $\varepsilon_{ij}$ is the mechanical strain, $e_{ijk}$ is the third-order tensor of piezoelectricity, $E_i$ is the electric field, $\mu_{ijkl}$ is the fourth-order total (including both direct and converse effects) flexoelectric tensor and $\kappa_{ij}$ is the second-order dielectric tensor.

The different stresses / electric displacements including the usual ($\hat{\sigma}_{ij}$ / $\widehat{D}_i$), higher-order ($\tilde{\sigma}_{ijk}$ / $\widetilde{D}_{ij}$) and physical ($\sigma_{ij}$ / $D_i$) ones are then defined through the following relations

$$\hat{\sigma}_{ij} = \frac{\partial \mathcal{H}}{\partial \varepsilon_{ij}} \quad \text{and} \quad \widehat{D}_i = -\frac{\partial \mathcal{H}}{\partial E_i} \tag{2}$$

$$\tilde{\sigma}_{ijk} = \frac{\partial \mathcal{H}}{\partial \varepsilon_{ij,k}} \quad \text{and} \quad \widetilde{D}_{ij} = -\frac{\partial \mathcal{H}}{\partial E_{i,j}} \tag{3}$$

$$\sigma_{ij} = \hat{\sigma}_{ij} - \tilde{\sigma}_{ijk,k} \quad \text{and} \quad D_i = \widehat{D}_i - \widetilde{D}_{ij,j} \tag{4}$$

thus

$$\sigma_{ij} = \hat{\sigma}_{ij} - \tilde{\sigma}_{ijk,k} = C_{ijkl} \varepsilon_{kl} - e_{kij} E_k + \mu_{lijk} E_{l,k} \tag{5}$$

$$D_i = \widehat{D}_i - \widetilde{D}_{ij,j} = e_{ikl} \varepsilon_{kl} + \kappa_{ij} E_j + \mu_{ijkl} \varepsilon_{jk,l} \tag{6}$$

which are the governing equations of the flexoelectricity. By imposing boundary conditions and integration over the domain, $\Omega$, the total electrical enthalpy is

$$H = \frac{1}{2} \int_\Omega \left( \hat{\sigma}_{ij} \varepsilon_{ij} + \tilde{\sigma}_{ijk} \varepsilon_{ij,k} - \widehat{D}_i E_i \right) d\Omega \tag{7}$$

Using Hamilton's principle, we finally have

$$\int_\Omega \left( C_{ijkl}\delta\varepsilon_{ij}\varepsilon_{kl} - e_{kij}E_k\delta\varepsilon_{ij} - \mu_{lijk}E_l\delta\varepsilon_{ij,k} - \kappa_{ij}\delta E_i E_j - e_{ikl}\delta E_i \varepsilon_{kl} - \mu_{ijkl}\delta E_i \varepsilon_{jk,l} \right) d\Omega$$

$$- \int_{\Gamma_t} \bar{t}_i \, \delta u_i dS + \int_{\Gamma_D} \varpi \, \delta\theta dS = 0 \tag{8}$$

which is the weak form of the governing equations of the flexoelectricity. In Eq. (8) $u_i$ is the mechanical displacements, $\theta$ is the electric potential, $\bar{t}_i$ is the prescribed mechanical tractions and $\varpi$ is the surface charge density. $\Gamma_t$ and $\Gamma_D$ are boundaries of $\Omega$ corresponding to mechanical tractions and electric displacements, respectively.

Using B-spline basis functions, $\boldsymbol{N}_u$ and $\boldsymbol{N}_\theta$, we approximate $\boldsymbol{u}$ and $\boldsymbol{\theta}$ fields as

$$u_h(x,y) = \sum_{i=1}^{ncp}\sum_{j=1}^{mcp} N_{i,j}^{p,q}(\xi,\eta) u_{ij}^e = (\boldsymbol{N}_u)^T \boldsymbol{u}^e \tag{9.a}$$

$$\theta_h(x,y) = \sum_{i=1}^{ncp}\sum_{j=1}^{mcp} N_{i,j}^{p,q}(\xi,\eta) \, \theta_{ij}^e = (\boldsymbol{N}_\theta)^T \boldsymbol{\theta}^e \tag{9.b}$$

where the superscripts $e$, $u$ and $\theta$ denote nodal parameters at the mesh control points, mechanical and electrical fields, respectively.

The discrete system of Eq. (8) is eventually expressed as

$$\begin{bmatrix} \boldsymbol{A}_{UU} & \boldsymbol{A}_{U\theta} \\ \boldsymbol{A}_{\theta U} & \boldsymbol{A}_{\theta\theta} \end{bmatrix} \begin{bmatrix} \boldsymbol{U} \\ \boldsymbol{\theta} \end{bmatrix} = \begin{bmatrix} \boldsymbol{f}_U \\ \boldsymbol{f}_\theta \end{bmatrix} \tag{10}$$

where

$$\boldsymbol{A}_{UU} = \sum_e \int_{\Omega_e} (\boldsymbol{B}_u) \, \boldsymbol{C} \, (\boldsymbol{B}_u)^T \, d\Omega \tag{11.a}$$

$$\boldsymbol{A}_{U\theta} = \sum_e \int_{\Omega_e} [(\boldsymbol{B}_u) \, \boldsymbol{e} \, (\boldsymbol{B}_\theta)^T + (\boldsymbol{H}_u) \, \boldsymbol{\mu}^T \, (\boldsymbol{B}_\theta)^T] d\Omega \tag{11.b}$$

$$\boldsymbol{A}_{\theta U} = \sum_e \int_{\Omega_e} [(\boldsymbol{B}_\theta) \, \boldsymbol{e}^T (\boldsymbol{B}_u)^T + (\boldsymbol{B}_\theta) \, \boldsymbol{\mu} \, (\boldsymbol{H}_u)^T] d\Omega \tag{11.c}$$

$$\boldsymbol{A}_{\theta\theta} = -\sum_e \int_{\Omega_e} (\boldsymbol{B}_\theta) \, \boldsymbol{\kappa} \, (\boldsymbol{B}_\theta)^T \, d\Omega \tag{11.d}$$

$$\boldsymbol{f}_U = \sum_e \int_{\Gamma_{t_e}} \boldsymbol{N}_u^T \boldsymbol{t}_\Gamma \, ds \tag{11.e}$$

$$\boldsymbol{f}_\theta = -\sum_e \int_{\Gamma_{D_e}} \boldsymbol{N}_\theta^T \varpi \, ds \tag{11.f}$$

In Eqs. (11.a-f), the subscript, $e$, in $\Omega_e$, $\Gamma_{t_e}$ and $\Gamma_{D_e}$ denotes the $e^{th}$ finite element where $\Omega = \bigcup_e \Omega_e$. Moreover, $\boldsymbol{B}_u$, $\boldsymbol{B}_\theta$ contain the spatial derivatives of the B-spline basis functions. The second derivatives of the basis functions, $\boldsymbol{H}_u$, are obtained by Eq. (12).

$$\boldsymbol{B}_u = \begin{bmatrix} \frac{\partial N_1}{\partial x} & 0 & \frac{\partial N_1}{\partial y} \\ \frac{\partial N_2}{\partial x} & 0 & \frac{\partial N_2}{\partial y} \\ \vdots & \vdots & \vdots \\ \frac{\partial N_{ncp}}{\partial x} & 0 & \frac{\partial N_{ncp}}{\partial y} \\ 0 & \frac{\partial N_1}{\partial y} & \frac{\partial N_1}{\partial x} \\ 0 & \frac{\partial N_2}{\partial y} & \frac{\partial N_2}{\partial x} \\ \vdots & \vdots & \vdots \\ 0 & \frac{\partial N_{ncp}}{\partial y} & \frac{\partial N_{ncp}}{\partial x} \end{bmatrix}, \boldsymbol{B}_\theta = \begin{bmatrix} \frac{\partial N_1}{\partial x} & \frac{\partial N_1}{\partial y} \\ \vdots & \vdots \\ \frac{\partial N_{ncp}}{\partial x} & \frac{\partial N_{ncp}}{\partial y} \end{bmatrix},$$

$$\boldsymbol{H}_u = \begin{bmatrix} \frac{\partial^2 N_1}{\partial x^2} & 0 & \frac{\partial^2 N_1}{\partial y \partial x} & \frac{\partial^2 N_1}{\partial x \partial y} & 0 & \frac{\partial^2 N_1}{\partial y^2} \\ \frac{\partial^2 N_2}{\partial x^2} & 0 & \frac{\partial^2 N_2}{\partial y \partial x} & \frac{\partial^2 N_2}{\partial x \partial y} & 0 & \frac{\partial^2 N_2}{\partial y^2} \\ \vdots & \vdots & \vdots & \vdots & \vdots & \vdots \\ \frac{\partial^2 N_{ncp}}{\partial x^2} & 0 & \frac{\partial^2 N_{ncp}}{\partial y \partial x} & \frac{\partial^2 N_{ncp}}{\partial x \partial y} & 0 & \frac{\partial^2 N_{ncp}}{\partial y^2} \\ 0 & \frac{\partial^2 N_1}{\partial y \partial x} & \frac{\partial^2 N_1}{\partial x^2} & 0 & \frac{\partial^2 N_1}{\partial y^2} & \frac{\partial^2 N_1}{\partial x \partial y} \\ 0 & \frac{\partial^2 N_2}{\partial y \partial x} & \frac{\partial^2 N_2}{\partial x^2} & 0 & \frac{\partial^2 N_2}{\partial y^2} & \frac{\partial^2 N_2}{\partial x \partial y} \\ \vdots & \vdots & \vdots & \vdots & \vdots & \vdots \\ 0 & \frac{\partial^2 N_{ncp}}{\partial y \partial x} & \frac{\partial^2 N_{ncp}}{\partial x^2} & 0 & \frac{\partial^2 N_{ncp}}{\partial y^2} & \frac{\partial^2 N_{ncp}}{\partial x \partial y} \end{bmatrix} \quad (12)$$

Moreover, $\boldsymbol{C}$, $\boldsymbol{\kappa}$, $\boldsymbol{e}$ and $\boldsymbol{\mu}$ can be written in matrix form as

$$\boldsymbol{C} = \left(\frac{Y}{(1+\nu)(1-2\nu)}\right) \begin{bmatrix} 1-\nu & \nu & 0 \\ \nu & 1-\nu & 0 \\ 0 & 0 & \left(\frac{1}{2}-\nu\right) \end{bmatrix} \quad (13.\text{a})$$

$$\boldsymbol{\kappa} = \begin{bmatrix} \kappa_{11} & 0 \\ 0 & \kappa_{33} \end{bmatrix} \quad (13.\text{b})$$

$$\boldsymbol{e}^{\mathrm{T}} = \begin{bmatrix} 0 & 0 & e_{15} \\ e_{31} & e_{33} & 0 \end{bmatrix} \quad (13.\text{c})$$

$$\boldsymbol{\mu} = \begin{bmatrix} \mu_{11} & \mu_{12} & 0 & 0 & 0 & \mu_{44} \\ 0 & 0 & \mu_{44} & \mu_{12} & \mu_{11} & 0 \end{bmatrix} \quad (13.\text{d})$$

where $\nu$ denotes Poisson's ratio and $Y$ is the Young's modulus.

## 3. Level Set Method (LSM) and optimization problem

### 3.1. LSM

Assume $\Omega_i \subset D \subset \mathbb{R}^d$ ($d = 2 \text{ or } 3$) where $D$ is the entire structural domain including all admissible shapes, $\Omega_i$. A single level set function $\Phi_i(x)$ is then defined as

$$\Phi_i(x): \begin{cases} \text{Phase 1: } \Phi_i(x) > 0 & \forall x \in \Omega_i \setminus \partial\Omega_i \\ \text{Boundary: } \Phi_i(x) = 0 & \forall x \in \partial\Omega_i \cap D \\ \text{Phase 2: } \Phi_i(x) < 0 & \forall x \in D \setminus \Omega_i \end{cases} \quad (14)$$

as schematically shown in Fig. (1.a). We use B-spline basis functions, $N_{i,j}^{p,q}$, to define $\Phi_i(x)$ according to

$$\Phi_i(x,y) = \sum_{i=1}^{ncp} \sum_{j=1}^{mcp} N_{i,j}^{p,q}(\xi,\eta)\, \varphi_{i,j} \quad (15)$$

where $ncp$, $mcp$ are the number of basis functions in the orthogonal directions and $\varphi_{i,j}$ denotes corresponding nodal values of the LS. As shown in Fig. (1.b), the zero iso-surface of $\Phi_i(x)$ implicitly represents the design boundary $\Gamma_i(x)$.

The level set function is dynamically updated at each time step by solving the Hamilton-Jacobi (H-J) partial differential equation

$$\frac{\partial \Phi_i}{\partial t} + V_i^n |\nabla \Phi_i| = 0 \quad (16)$$

in which $V_i^n = \boldsymbol{V}_i \cdot \boldsymbol{n}_i$ is the normal component of the velocity vector ($\boldsymbol{V}_i = (\frac{dx}{dt})_i$) and $\boldsymbol{n}_i = \frac{\nabla \Phi_i}{|\nabla \Phi_i|}$ is the unit outward normal to the boundary $\Gamma_i$. The field $V_i^n$ determines geometric motion of the boundary $\Gamma_i$ and is chosen based on the design sensitivity of the objective function. $\Phi_i$ is initiated as a signed distance function and the above H-J equation is solved by an explicit first-order upwind scheme [19].

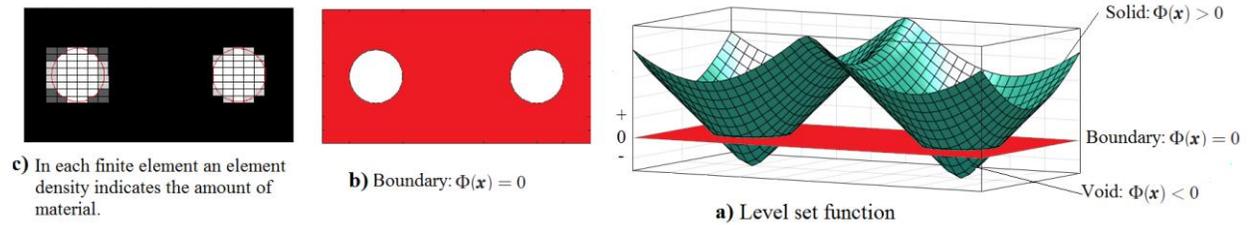

**Fig.1.** Level set function (**a**), boundary representation with level set (**b**) and density mapping technique (**c**)

Conventional partitioning of the whole material domain into $n$ phases, $\omega_1, \ldots, \omega_n$, (including the void phase) using $m = n - 1$ LS functions where each one represents a distinct material phase [20] introduces a range of computational challenges: 1) numerical difficulties to maintain the "partition conditions" $D = \bigcup_{k=1}^{n} \omega_k$ and $\omega_k \cap \omega_l = \emptyset$, $k \neq l$ and 2) complexity associated with a high number of level set functions. To remove these shortcomings, we follow [21] and adopt the vector LS approach [22] where a number of $m$ level set function partitions the design domain $D$ into $n = 2^m$ overlapping regions, $\omega_k$ $(k = 1, \ldots, n)$, obtained by different combinations of the zero-level sets $\Omega_i (i = 1, \ldots, m)$. In this scheme the interior regions of the zero-level sets of these functions $\Omega_i = \{x: \Phi_i(x) > 0\}$ can overlap. Thus, each point $x \in D$ belongs to one and only one material phase which essentially satisfies the partition conditions [21].

We will focus on examining flexoelectric composites using up to two level set functions. Fig. (2) illustrates four material phases defined by two level-set functions $\Phi_1$ and $\Phi_2$. For the case of three phases (including the void phase), $\Phi_1$ determines the solid and the void phases while $\Phi_2$ distinguishes different solid material phases.

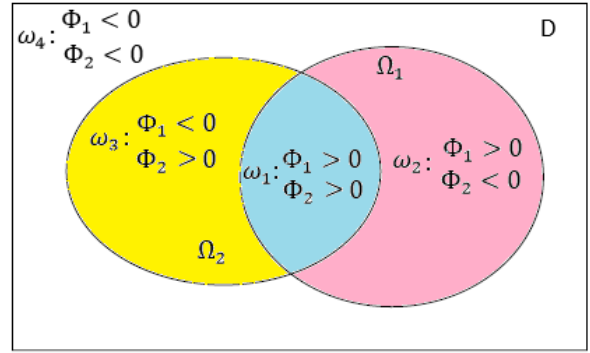

**Fig.2.** Four material phases $\omega_1, \omega_2, \omega_3, \omega_4$ are represented by two level-set functions $\Phi_1$ and $\Phi_2$

We consider the vector level-set function $\boldsymbol{\Phi} = [\Phi_1 \ \Phi_2 \ \ldots \ \Phi_m]$ and the vector Heaviside function $\widetilde{\boldsymbol{H}}(\boldsymbol{\Phi}) = [\widetilde{H}(\Phi_1) \ \widetilde{H}(\Phi_2) \ \ldots \ \widetilde{H}(\Phi_m)]$ where $\widetilde{H}(\Phi_i)$ is a smooth approximation of the Heaviside function defined by

$$\widetilde{H}(\Phi_i) = \begin{cases} 0 & \text{for } \Phi_i < -\Delta \\ -\frac{1}{4}\left(\frac{\Phi_i}{\Delta}\right)^3 + \frac{3}{4}\left(\frac{\Phi_i}{\Delta}\right) + \frac{1}{2} & \text{for } -\Delta \leq \Phi_i \leq \Delta \\ 1 & \text{for } \Delta < \Phi_i \end{cases} \quad (17)$$

where $\Delta$ is the width of numerical approximation.

Using point-wise mapping to control an element-wise constant phase density distribution (as represented in Fig. (1.c) for a single material phase), we define

$$\text{for two phases:} \begin{cases} \rho_1 = \widetilde{H}(\Phi_1(X_e)) & Active\ 1 \\ \rho_2 = 1 - \rho_1 = 1 - \widetilde{H}(\Phi_1(X_e)) & Active\ 2 \end{cases} \quad (18.a)$$

,

$$\text{for three phases:} \begin{cases} \rho_1 = \widetilde{H}(\Phi_1(X_e)) & solid \\ \rho_2 = \widetilde{H}(\Phi_1(X_e)) \times \widetilde{H}(\Phi_2(X_e)) & Active\ 1\ (flexo) \\ \rho_3 = \rho_1 - \rho_2 = \widetilde{H}(\Phi_1(X_e)) \times (1 - \widetilde{H}(\Phi_2(X_e))) & elastic \\ \rho_4 = 1 - \rho_1 = 1 - \widetilde{H}(\Phi_1(X_e)) & void \end{cases}$$

(18.b)

and

$$\text{for four phases:} \begin{cases} \rho_1 = \widetilde{H}(\Phi_1(X_e)) \times \widetilde{H}(\Phi_2(X_e)) & Active\ 1 \\ \rho_2 = \widetilde{H}(\Phi_1(X_e)) \times (1 - \widetilde{H}(\Phi_2(X_e))) & elastic \\ \rho_3 = (1 - \widetilde{H}(\Phi_1(X_e))) \times \widetilde{H}(\Phi_2(X_e)) & Active\ 2\ or\ 3 \\ \rho_4 = (1 - \widetilde{H}(\Phi_1(X_e))) \times (1 - \widetilde{H}(\Phi_2(X_e))) & void \end{cases} \quad (18.c)$$

where $0 \leq \rho_k \leq 1$ and $X_e$ is the center of a finite element $e$. These element densities are embedded in the electromechanical problem to obtain effective material properties

$$\boldsymbol{M}_{eff}(x, y) = \sum_{k=1}^{2} \rho_k(\boldsymbol{\varphi}) \, \boldsymbol{M}_k^0 \quad \text{for two phases} \quad (19.a)$$

$$\boldsymbol{M}_{eff}(x, y) = \sum_{k=2}^{4} \rho_k(\boldsymbol{\varphi}) \, \boldsymbol{M}_k^0 \quad \text{for three phases} \quad (19.b)$$

$$\boldsymbol{M}_{eff}(x, y) = \sum_{k=1}^{4} \rho_k(\boldsymbol{\varphi}) \, \boldsymbol{M}_k^0 \quad \text{for four phases} \quad (19.c)$$

where Eqs. (13.a-d) define $\boldsymbol{M}_k^0 = \boldsymbol{C}_k^0, \boldsymbol{e}_k^0, \boldsymbol{\kappa}_k^0, \boldsymbol{\mu}_k^0$. Superscript 0 represents properties of the bulk materials. $\boldsymbol{C}^0$ and $\boldsymbol{\kappa}_k^0$ for the void phase contain appropriately small values to avoid singularity of the stiffness matrix.

Assuming $\boldsymbol{\rho} = [\rho_k]$ where $k = 1, \dots, n$, the volume integrals of some functional $f$ over a material domain can then be defined as

$$\int_\Omega f dV = \int_D f \widetilde{H}(\boldsymbol{\Phi}) dV \approx \int_D f \boldsymbol{\rho}(\boldsymbol{\varphi}) \, dV \quad (20)$$

where $\boldsymbol{\varphi}$ is a matrix containing all vectors of $\boldsymbol{\varphi}_i$ ($i = 1, \dots, m$). Each vector $\boldsymbol{\varphi}_i$ (associated with the LS function $\Phi_i$) contains related design variables, $\varphi_{i',j'}$ defined on the mesh of control points.

### 3.2. Optimization problem

The electromechanical coupling coefficient, $k^2$, is defined as

$$k^2 = \frac{w_{elec}}{w_{mech}} \qquad (21)$$

where $w_{elec}$ and $w_{mech}$ are the electrical and mechanical (or strain) energies, respectively. By extending $w_{elec}$ and $w_{mech}$ in Eq. (21) and defining the objective function, $J(\boldsymbol{u}(\boldsymbol{\varphi}), \boldsymbol{\theta}(\boldsymbol{\varphi}), \boldsymbol{\varphi})$, as the inverse of $k^2$ we have

$$J(\boldsymbol{u}(\boldsymbol{\varphi}), \boldsymbol{\theta}(\boldsymbol{\varphi}), \boldsymbol{\varphi}) = \frac{1}{k^2} = \frac{w_{mech}}{w_{elec}} = \frac{\frac{1}{2}\int_\Omega \boldsymbol{\varepsilon}^\mathrm{T} \boldsymbol{C}\, \boldsymbol{\varepsilon}\, d\Omega}{\frac{1}{2}\int_\Omega \boldsymbol{E}^\mathrm{T}\boldsymbol{\kappa}\, \boldsymbol{E}\, d\Omega} \qquad (22)$$

where $\boldsymbol{\varepsilon} = (\boldsymbol{B}_u)^\mathrm{T} \boldsymbol{u}^e$ and $\boldsymbol{E} = -(\boldsymbol{B}_\theta)^\mathrm{T} \boldsymbol{\theta}^e$. Eventually, in its general form the optimization problem can be summarized as Eq. (23) and Table-1:

$$\begin{cases} \text{Minimize: } J(\boldsymbol{u}(\boldsymbol{\varphi}), \boldsymbol{\theta}(\boldsymbol{\varphi}), \boldsymbol{\varphi}) \\ \text{Subjected to:} \\ V_k = \int_D \rho_k(\boldsymbol{\Phi})\, d\Omega = V_{k0} \qquad k = 1, \dots, n \\ \begin{bmatrix} \boldsymbol{A}_{UU} & \boldsymbol{A}_{U\theta} \\ \boldsymbol{A}_{\theta U} & \boldsymbol{A}_{\theta\theta} \end{bmatrix} \begin{bmatrix} \boldsymbol{U} \\ \boldsymbol{\theta} \end{bmatrix} = \begin{bmatrix} \boldsymbol{f}_U \\ \boldsymbol{f}_\theta \end{bmatrix} \end{cases} \qquad (23)$$

**Table-1** Summary of the optimization problem

| | | | |
|---|---|---|---|
| **Inputs** | Initial nodal values of the level set functions, $\varphi_{i,j}^{initial}$ | Material properties | Solver settings & parameters |
| **Design variables** | Nodal values of the level set functions, $\varphi_{i,j}$ | | |
| **Design constraints** | Volume of the material phases, $V_k$ where $k = 1, \dots, n$ | System of coupled governing equations | |
| **outputs** | Optimum distributions of material phases, $\varphi_{i,j}^{optimum}$ | | |

where $V_k$ is the total volume of the material phase $k$ in each optimization iteration and $V_{k0}$ is the corresponding given volume.

To satisfy the volume constraints, we use the augmented Lagrangian method combining the properties of the Lagrangian (the second term in Eq. (24.a)) and the quadratic penalty functions (the third term in Eq. (24.a)). It seeks the solution by replacing the original constrained problem by a sequence of unconstrained sub-problems through estimating explicit Lagrangian multipliers at each step to avoid the ill-conditioning that is inherent in the quadratic penalty function (see [23] for more details).

Following [23], we define

$$l = J + \sum_{k=1}^{k=n} \psi_k^j (V_k - V_{k0}) + \frac{1}{2\Lambda_k^j}(V_k - V_{k0})^2 \qquad (24.a)$$

$\psi_k^j$ and $\Lambda_k^j$ are parameters in $j^{th}$ iteration which are updated according to the following scheme

$$\psi_k^{j+1} = \psi_k^j + \frac{1}{\Lambda_k^j}(V_k - V_{k0}) \quad , \quad \Lambda_k^{j+1} = \alpha \Lambda_k^j \qquad (24.b)$$

where $\alpha \in (0,1)$ is a fixed parameter. $\psi_k^j$ and $\Lambda_k^j$ start with appropriately chosen initial values; then $\boldsymbol{\varphi}$ that approximately minimizes $l$ will be found. $\psi_k^j$ and $\Lambda_k^j$ are subsequently updated and the process is repeated until the solution converges.

The classical Lagrangian objective function is obtained by discarding the last term of Eq. (24.a). The normal velocity $V_i^n$ in Eq. (16) is chosen as a descent direction for the Lagrangian $l$ according to

$$V_i^n = -\frac{dl}{d\varphi_i} = -\frac{dJ}{d\varphi_i} - [\psi_k^j + \frac{1}{\Lambda_k^j}(V_k - V_{k0})]\frac{dV_k}{d\varphi_i} \qquad (i = 1, \ldots, m) \qquad (25)$$

where different terms of Eq. (25) are derived in Appendix A. The flowchart of the entire optimization process is presented in Fig.3.

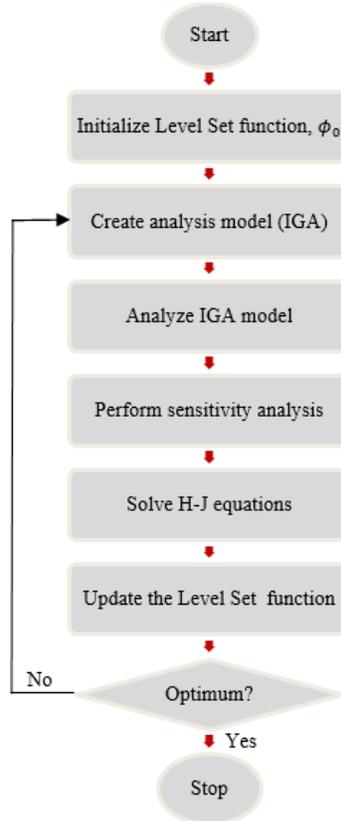

**Fig.3.** The flowchart of the optimization process

## 4. Numerical examples

We perform a suite of examples of multi-material beams with linear elastic material properties and under 2D plane strain conditions. In all following examples, we assume a $60 \times 15 \ \mu m$ cantilever beam discretized by $48 \times 12$ quadratic B-spline elements, unless otherwise specified. The beam is subjected to a downward point load of $100 \ \mu N$ at the top of the free edge while open circuit electrical boundary conditions are imposed as shown in Fig. (4.a). We investigate two, three and four phase composite beams. All models are discretized by quadratic B-spline elements (see Fig. (4.b)) where red dots represent control points (see [10] for more details).

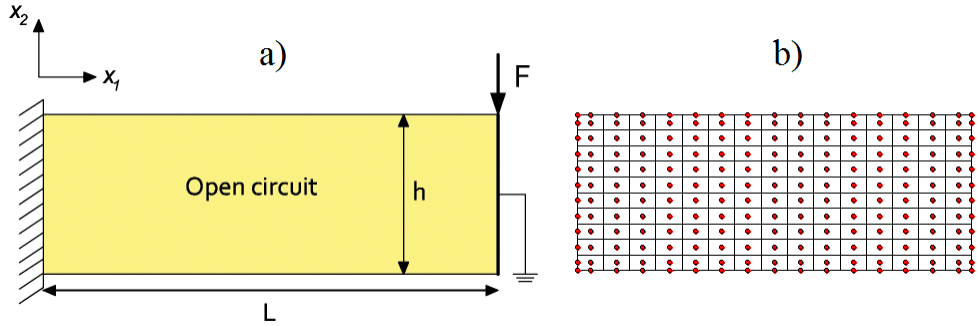

**Fig.4.** Loading and boundary conditions (**a**), discretization (**b**)

Table-2 includes material properties of the active (piezoelectric or flexoelectric), passive (elastic) and void phases. An active non-piezoelectric material experiences pure flexoelectricity and is obtained by setting $e_{ij} = 0$.

**Table-2** Properties of active 1 ($BaTiO_3$ [16]), active 2, active 3, passive and void phases

| Phase / color | $\upsilon$ | Y | $e_{31}$ | $\mu_{11}/\mu_{12}$ | $\kappa_{11}$ | $\kappa_{33}$ |
|---|---|---|---|---|---|---|
| Active 1 / blue | 0.37 | $100 \ GPa$ | $-4.4 \ C/m^2$ | $1 \ \mu C/m$ | $11 \ nC/Vm$ | $12.48 \ nC/Vm$ |
| Active 2 / red | 0.37 | $50 \ GPa$ | $-2.2 \ C/m^2$ | $0.5 \ \mu C/m$ | $5.5 \ nC/Vm$ | $6.24 \ nC/Vm$ |
| Active 3 / yellow | 0.37 | $100 \ GPa$ | $-4.4 \ C/m^2$ | 0 | $11 \ nC/Vm$ | $12.48 \ nC/Vm$ |
| Passive / green | 0.37 | $10 \ GPa$ | 0 | 0 | $0.02 \ nC/Vm$ | $0.02 \ nC/Vm$ |
| Void / white | 0.37 | $1 \ GPa$ | 0 | 0 | $0.0089 \ nC/Vm$ | $0.0089 \ nC/Vm$ |

$\upsilon$: $Poisson's \ ratio$, $\quad Y$: $Young's \ modulus$, $\quad e_{31}$: $piezoelectric \ constant$,
$\mu_{11}/\mu_{12}$: $flexoelectric \ constants$, $\quad \kappa_{11}/\kappa_{33}$: $dielectric \ constants$

### 4.1. Two phase composite

In this section, we assume the beam is made from: the non-piezoelectric (i.e. setting $e_{31} = 0$) Active 1 and the passive elastic phases (Case-1), and the non-piezoelectric Active 1 and Active 2 phases (Case-2) according to Table-2. For both cases, the electromechanical coupling coefficient, $k^2$, is measured for various compositions of constituent phases, while the normalized electromechanical coupling coefficient ($k_n^2$) is obtained by normalizing the cases by the electromechanical coupling coefficient of the beam with 100% Active 1 material. Fig. (5) belongs to the Case-1 and it is observable that, by combining the passive and the active phases a higher $k_n^2$ than the single-phase counterpart can be obtained; however, there is a point where the result is optimal. In fact, more soft passive material on the one hand increases $w_{mech}$, which subsequently decreases $k^2$, but on the other hand, it produces higher strain and strain gradients, which gives rise to higher $w_{elec}$ and $k^2$. Thus, in the optimal material combination there is a tradeoff between these two conflicting effects.

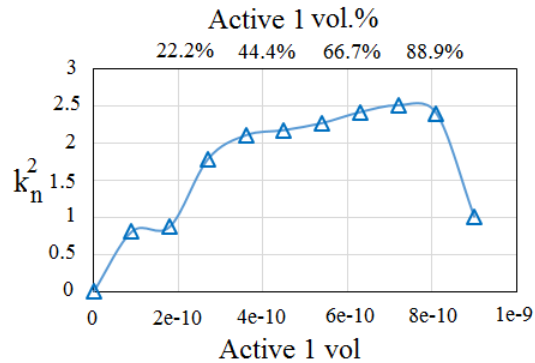

**Fig.5.** $k_n^2$ versus volume fraction of Active 1 for Case-1

Fig. (6) represents the results for Case-2 while the optimized topologies are presented as well. One can observe that any combination of the Active 1 and Active 2 lead to the higher $k^2$ than either the single-phase Active 1 or Active 2 counterparts.

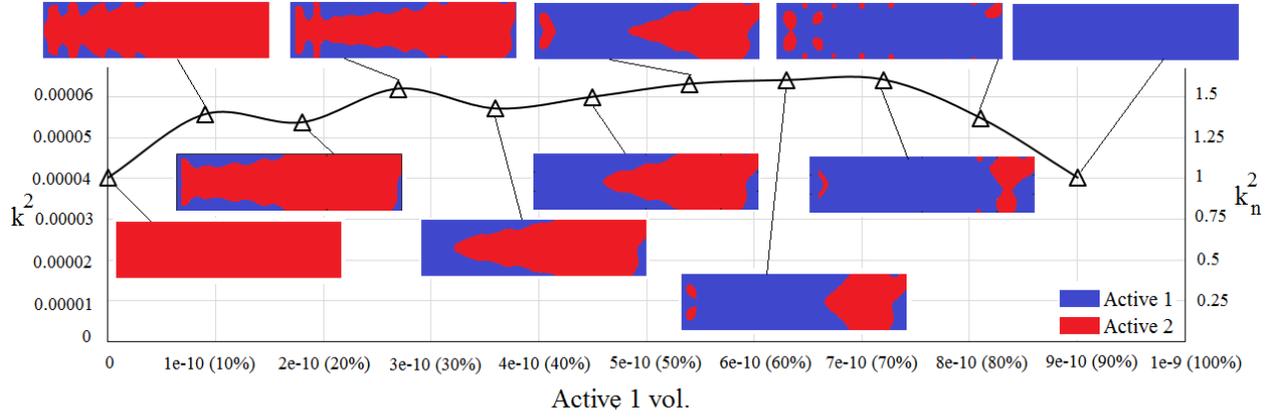

**Fig.6.** $k^2$ and $k_n^2$ versus volume fraction of Active 1 for Case-2

### 4.2. Three phase composite

Let us assume the non-piezoelectric active, passive and void phases (Active 1, Passive and Void in Table-2). Fig. (7.a-e) include the optimal topologies. As mentioned in Table-2, the flexoelectric phase ($\rho_2$) is shown in blue, elastic ($\rho_3$) in green and void ($\rho_4$) in white colors. The solid phase in Fig. (7.a) only includes the flexoelectric phase (zero elastic phase); while in Fig. (7.b) $V_{\text{flexo}} = \int_D \rho_2 \, d\Omega = 0.56 \times V_0$, $V_{\text{elastic}} = \int_D \rho_3 \, d\Omega = 0.14 \times V_0$ and $V_{\text{void}} = \int_D \rho_4 \, d\Omega = 0.3 \times V_0$ are considered as volume constraints. $\rho_2$, $\rho_3$ and $\rho_4$ are calculated according to Eq. (18). We write these constraints in compact form as $[V_{flexo}, V_{elastic}, V_{void}] = [0.56, 0.14, 0.3] \times V_0$ in which $V_0 = L_x \times L_y$. We also set $[V_{flexo}, V_{elastic}, V_{void}] = [0.42, 0.28, 0.3] \times V_0$ in Fig. (7.c), $[V_{flexo}, V_{elastic}, V_{void}] = [0.28, 0.42, 0.3] \times V_0$ in Fig. (7.d) and $[V_{flexo}, V_{elastic}, V_{void}] = [0.14, 0.56, 0.3] \times V_0$ in Fig. (7.e).

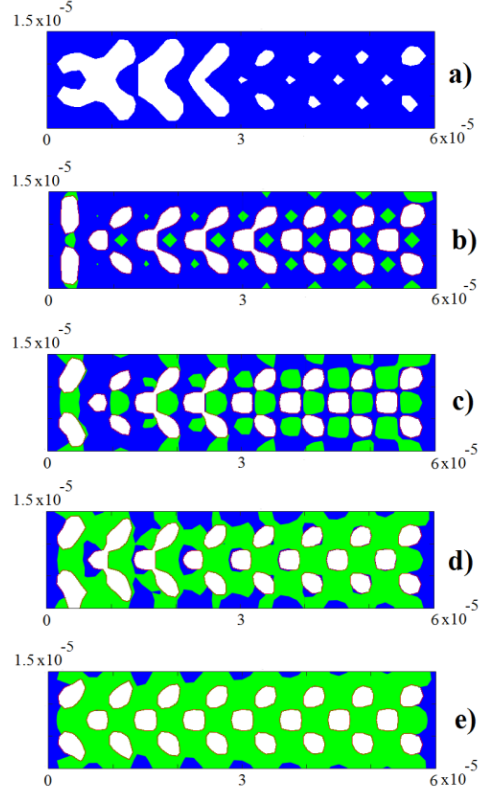

**Fig.7.** The optimal topologies for the flexoelectric beam considering $V_{flexo} = 0.7V_0$ (**a**), $V_{flexo} = 0.56V_0$ (**b**), $V_{flexo} = 0.42V_0$ (**c**), $V_{flexo} = 0.28V_0$ (**d**) and $V_{flexo} = 0.14V_0$ (**e**) where $V_0 = L_x \times L_y$. In all insets $V_{void} = 0.3V_0$. The flexoelectric phase is shown in blue, elastic in green and void in white colors.

Browsing Fig. (7) from the top towards the bottom, one can visually find that the elastic (green) phase increases, the flexoelectric (blue) phase decreases and the void (white) remains constant. Furthermore, because of the larger strain gradients around the perimeter, the flexoelectric phase concentrates on the outside (perimeter) of the beam, whereas the elastic material is in the interior. A rigorous scrutiny of the volume constraints fulfillment as well as the objective function minimization is presented in Fig. (8). The graphs belong to Fig. (7.e) and illustrate how the volumes and the objective function converge precisely and smoothly towards the specified or minimum values.

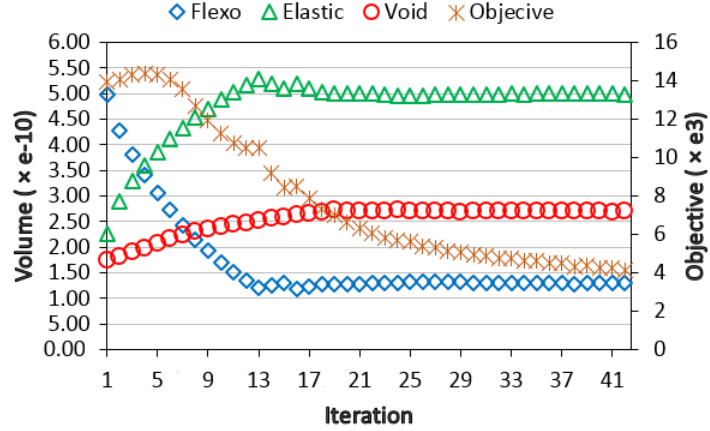

**Fig.8.** Phases volumes and objective function versus iteration for $[V_{flexo}, V_{elastic}, V_{void}] = [0.14, 0.56, 0.3]V_0$ (corresponding to Fig.7(e))

To find how the elastic phase impacts the efficiency of the device, $k^2$ is measured for each inset of Fig. (7) and the normalized results, $k_n^2$, (by the solid beam with 100% flexoelectric phase) are presented in Fig. (9.a). The void phase is constant ($0.3V_0$) in all cases and the solid material can have different combinations of the flexoelectric and elastic phases. For the bulk elastic structure, $k^2$ is zero since there is no active material. When 14% flexoelectric phase is added $k^2$ becomes $\approx 0.00022$ ($k_n^2 = 5.57$) and for 28% flexoelectric, $k^2$ is $\approx 0.00037$ ($k_n^2 = 9.14$). Interestingly, by increasing the flexoelectric phase to 0.42% not only does $k^2$ not increase but it instead decreases to the value of $\approx 0.00033$ ($k_n^2 = 8.1$). Further increasing the flexoelectric phase yields further reduction in $k^2$ i.e. $k^2 \approx 0.00016$ ($k_n^2 = 3.98$) for the flexoelectric device with 70% flexoelectric and 30% void phases.

We repeat the problem by measuring $k_n^2$ of the beam with the same length and the aspect ratio of 6. The similar trend is observed as shown in Fig. (9.b). We observe that by combining the passive and the active phases a higher electromechanical coefficient than the single-phase counterpart can be obtained.

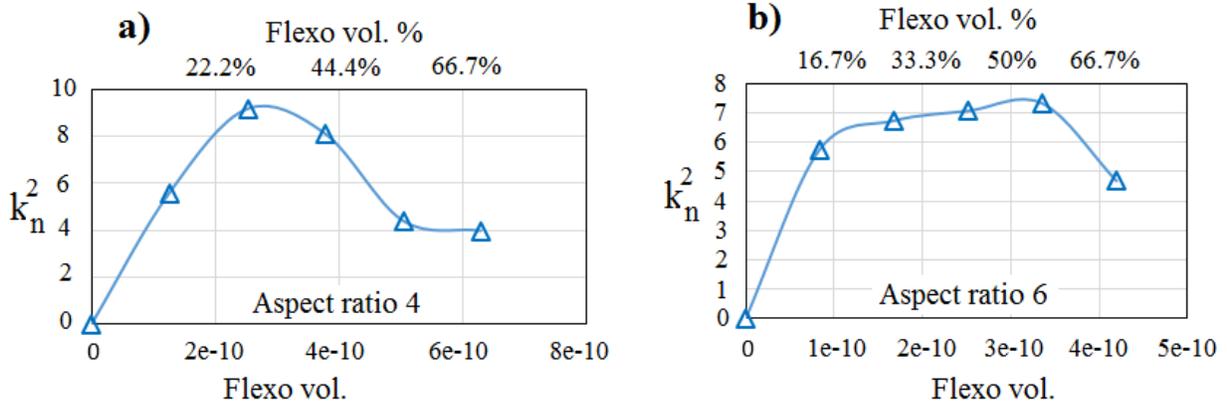

**Fig.9.** The normalized electromechanical coupling coefficient, $k_n^2$, versus volume fraction of the flexoelectric phase for the beam with aspect ratio of 4 (**a**) and 6 (**b**). For all cases, the void phase is kept constant as $0.3V_0$ where $V_0 = L_x \times L_y$. The length of the beam is 60 $\mu m$.

Fig. (10) includes the optimal topologies for different beam aspect ratios of 4, 6 and 8 considering $[V_{flexo}, V_{elastic}, V_{void}] = [0.28, 0.42, 0.3] \times V_0$. The results are presented in Table-3. For comparing results, it should be noted that the flexoelectric size effect and the volume ratio of the flexoelectric material are contradictory. The former causes the highest $k^2$ for the beam with the aspect ratio of 8 though there is less active material to generate electricity in comparison with the smaller aspect ratio beams. The latter makes $k^2$ for the beam with the aspect ratio of 6 be smaller than $k^2$ when the beam aspect ratio is 4. It is obvious that for the solid beams, larger aspect ratio leads to larger $k^2$ (see [10]).

**Table-3** $k^2$ and $k_n^2$ for different beam aspect ratios

| Aspect ratio | 4 | 6 | 8 |
|---|---|---|---|
| $V_{flexo}$ | $2.52e-10$ | $1.68e-10$ | $1.26e-10$ |
| $k^2$ | 0.00037 | 0.00029 | 0.00072 |
| $k_n^2$ | 9.14 | 6.73 | 13.63 |

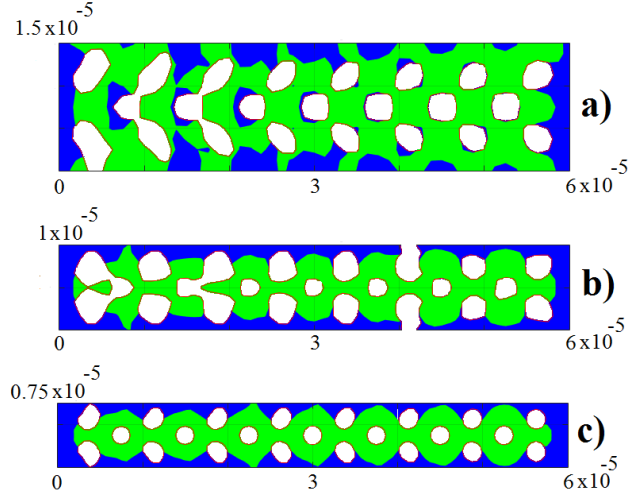

**Fig.10.** The optimal topologies of the beam with aspect ratios 4 (**a**), 6 (**b**) and 8 (**c**). In all examples: $V_{flexo} = 0.28V_0$ and $V_{void} = 0.3V_0$ where $V_0 = L_x \times L_y$. The length of the beam is $60\ \mu m$.

### 4.3. Four phase composite

Here, we consider the beam made from four phases, as presented in Table-2, through two cases: Active 1 and Active 2 phases are considered as non-piezoelectric ($e_{31} = 0$) materials (Case-1) and Active 1 as a non-piezoelectric material and Active 3 as a pure piezoelectric ($\mu_{11}/\mu_{12} = 0$) material without any flexoelectric properties (Case-2). In both cases, there are also void and elastic phases and $[V_{active1}, V_{elastic}, V_{active2}, V_{void}] = [0.21, 0.28, 0.21, 0.3] \times V_0$ are set as volume constraints.

Fig. (11) and Fig. (12) show optimal topologies for Case-1 and Case-2, respectively. For each case the history of the objective function and volume constraints are presented separately.

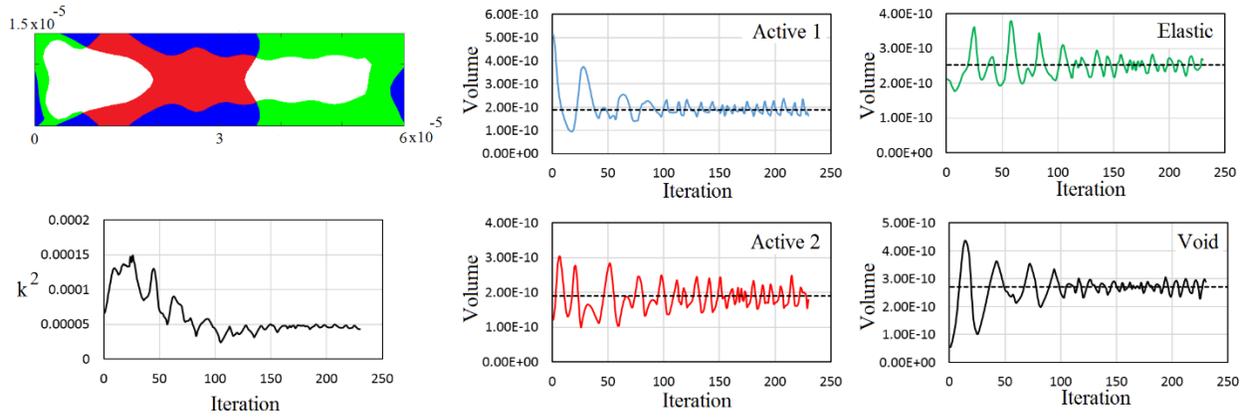

**Fig.11.** The optimal topology for Case-1 composed of Active 1 (blue), Active 2 (red), elastic (green) and hole (white) phases. $[V_{active1}, V_{elastic}, V_{active2}, V_{void}] = [0.21, 0.28, 0.21, 0.3] \times V_0$ where $V_0 = L_x \times L_y$ are set as four equality design constraints. The length of the beam is $60~\mu m$ and its aspect ratio is 4.

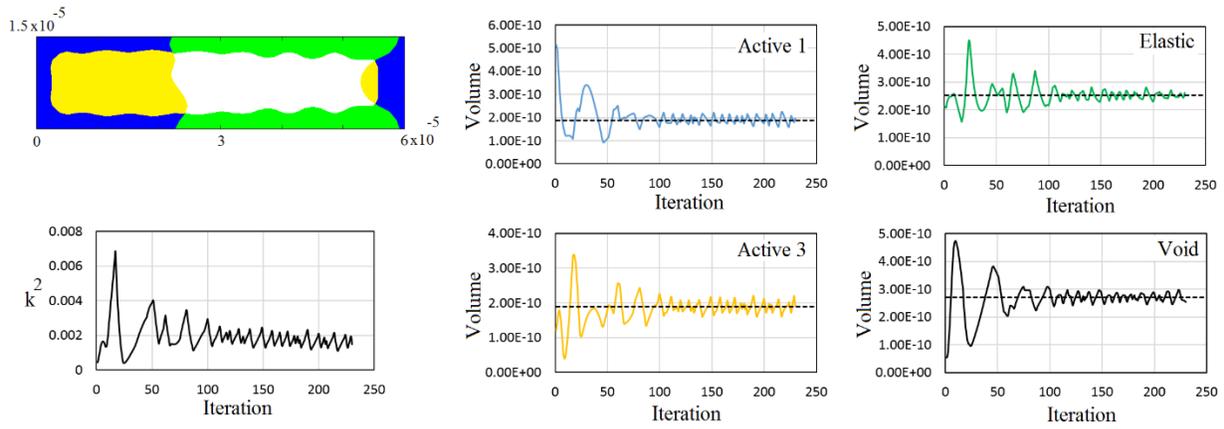

**Fig.12.** The optimal topology for Case-2 composed of Active 1 (blue), Active 3 (yellow), elastic (green) and hole (white) phases. $[V_{active1}, V_{elastic}, V_{active2}, V_{void}] = [0.21, 0.28, 0.21, 0.3] \times V_0$ where $V_0 = L_x \times L_y$ are set as four equality design constraints. The length of the beam is $60~\mu m$ and its aspect ratio is 4.

## 5. Concluding remarks

The B-spline elements which were successfully implemented to model flexoelectric effect in dielectric materials are combined with the vector level set technique, with the goal of enhancing the electromechanical performance of multi-phase micro and nano sensors and actuators made from different active (flexoelectric and piezoelectric) and passive (elastic) materials.

The numerical examples show the capabilities of the model to design two, three and four phase micro sensors with the optimal electromechanical coupling coefficient defined by $k^2 = \frac{w_{elec}}{w_{mech}}$ where $w_{elec}$ and $w_{mech}$ are the electrical and mechanical energies, respectively. For the two phase composite made from the active and passive phases, our results show that at the optimal volume fractions of constituents, the normalized electromechanical coupling coefficient ($k_n^2$) is 2.5 times larger than what that obtained from a beam made purely from the active material. For the three phase composite case (made from active and passive materials as well as holes), $k_n^2$ is increased by a factor of 9. The results demonstrate the competing effects of increasing volume fraction of the soft passive material in the composite, which on the one hand decreases $k^2$ by increasing $w_{mech}$, and on the other hand, increases $k^2$ by increasing $w_{elec}$ by producing higher strain gradients. Thus, in the optimal materials combination there is a tradeoff between these two competing effects.

Future work will focus on studies on numerical stability, updating procedure, geometry mapping and regularization. One crucial aspect of the method is the determination of the Lagrange multipliers to minimize the objective function while the multiple equality volume constraints are also precisely fulfilled. It is possible that an optimality criteria method would better treat this kind of constraint by means of the move limit and the damping factor; however, the LS function is susceptible to becoming too flat or too steep, both of which may give rise to convergence issues.

**Acknowledgments:**

Hamid Ghasemi and Timon Rabczuk gratefully acknowledge the financial support by European Research Council for COMBAT project (Grant number 615132). Harold Park acknowledges the support of the Mechanical Engineering department at Boston University.

**Appendix A: Sensitivity analysis**

The coupled system of equations in a single global residual form is expressed as

$$\mathfrak{R}(\mathfrak{U}) = \begin{bmatrix} R_1^*(u,\theta) \\ R_2^*(u,\theta) \end{bmatrix} = \mathfrak{R}(\mathfrak{U}(\varphi),\varphi) = 0 \tag{A1}$$

where $R_1^*$ and $R_2^*$ are residuals that must be simultaneously satisfied; $\mathfrak{U} = \begin{bmatrix} u \\ \theta \end{bmatrix}$ where $u$ and $\theta$ are solution (i.e. displacement and electric potential) fields. The objective function then takes the form $J(\mathfrak{U}(\boldsymbol{\varphi}), \boldsymbol{\varphi})$. We calculate the sensitivity of the objective function, $J(\mathfrak{U}(\boldsymbol{\varphi}), \boldsymbol{\varphi})$, and volume constraints, $V_{\text{solid}}$ and $V_{\text{flexo}}$, in Eq. (25) with respect to $\boldsymbol{\varphi}_1$ and $\boldsymbol{\varphi}_2$. Using the chain-rule we have

$$\frac{dJ}{d\varphi_i} = \frac{\partial J}{\partial \mathfrak{u}}\frac{\partial \mathfrak{u}}{\partial \varphi_i} + \frac{\partial J}{\partial \varphi_i} = \frac{\partial J}{\partial \mathfrak{u}}\left[-\left(\frac{\partial \mathfrak{R}}{\partial \mathfrak{u}}\right)^{-T}\frac{\partial \mathfrak{R}}{\partial \varphi_i}\right] + \frac{\partial J}{\partial \varphi_i} = (\boldsymbol{\lambda})^T \frac{\partial \mathfrak{R}}{\partial \varphi_i} + \frac{\partial J}{\partial \varphi_i} \quad (i = 1, \dots, m) \quad \text{(A2)}$$

where $\boldsymbol{\lambda} = -\frac{\partial J}{\partial \mathfrak{u}}\left(\frac{\partial \mathfrak{R}}{\partial \mathfrak{u}}\right)^{-T}$ and the term inside the brackets is obtained by differentiating Eq. (A1) as

$$\left(\frac{\partial \mathfrak{R}}{\partial \mathfrak{u}}\right)^T \frac{\partial \mathfrak{u}}{\partial \varphi_i} + \frac{\partial \mathfrak{R}}{\partial \varphi_i} = \mathbf{0} \quad \text{(A3)}$$

By substituting $\frac{\partial \mathfrak{R}}{\partial \mathfrak{u}} = \boldsymbol{K}_{\text{total}}$ into the definition of $\boldsymbol{\lambda}$ we obtain

$$\boldsymbol{K}_{\text{total}}\boldsymbol{\lambda} = -\frac{\partial J}{\partial \mathfrak{u}} = -\frac{1}{w_{elec}}\left(\int_D \boldsymbol{B}_u \boldsymbol{C} \boldsymbol{B}_u^T \boldsymbol{u}\, d\Omega\right) + \frac{w_{\text{mech}}}{w_{elec}^2}\left(\int_D \boldsymbol{B}_\theta \boldsymbol{\kappa} \boldsymbol{B}_\theta^T \boldsymbol{\theta}\, d\Omega\right) \quad \text{(A4)}$$

Having obtained $\boldsymbol{\lambda}$, one can write

$$(\boldsymbol{\lambda})^T \frac{\partial \mathfrak{R}}{\partial \varphi_i} = \begin{bmatrix} A'_{UU} & A'_{U\theta} \\ A'_{\theta U} & A'_{\theta\theta} \end{bmatrix}[\boldsymbol{\lambda}] \quad \text{(A5)}$$

where

$$A'_{UU} = \sum_e \int_{D_e} \boldsymbol{u}^T \left(\boldsymbol{B}_u \frac{\partial \boldsymbol{C}}{\partial \varphi_i} \boldsymbol{B}_u^T\right) d\Omega \quad \text{(A6)}$$

$$A'_{U\theta} = \sum_e \int_{D_e} \boldsymbol{u}^T \left(\boldsymbol{B}_u \frac{\partial \boldsymbol{e}}{\partial \varphi_i} \boldsymbol{B}_\theta^T + \boldsymbol{H}_u \frac{\partial \boldsymbol{\mu}^T}{\partial \varphi_i} \boldsymbol{B}_\theta^T\right) d\Omega \quad \text{(A7)}$$

$$A'_{\theta U} = \sum_e \int_{D_e} \boldsymbol{\theta}^T \left(\boldsymbol{B}_\theta \frac{\partial \boldsymbol{e}^T}{\partial \varphi_i} \boldsymbol{B}_u^T + \boldsymbol{B}_\theta \frac{\partial \boldsymbol{\mu}}{\partial \varphi_i} \boldsymbol{H}_u^T\right) d\Omega \quad \text{(A8)}$$

$$A'_{\theta\theta} = -\sum_e \int_{D_e} \boldsymbol{\theta}^T \left(\boldsymbol{B}_\theta \frac{\partial \boldsymbol{\kappa}}{\partial \varphi_i} \boldsymbol{B}_\theta^T\right) d\Omega \quad \text{(A9)}$$

and

$$\frac{\partial \boldsymbol{M}}{\partial \varphi_i} = \sum_{k=1}^{4} \frac{\partial \rho_k}{\partial \varphi_i} \boldsymbol{M}_k^0 \quad \text{with } i = 1, \dots, m \text{ and } \boldsymbol{M}_k^0 = \boldsymbol{C}_k^0, \boldsymbol{e}_k^0, \boldsymbol{\kappa}_k^0, \boldsymbol{\mu}_k^0 \quad \text{(A10)}$$

where $m$ is the number of level set functions and $\boldsymbol{C}_k^0, \boldsymbol{e}_k^0, \boldsymbol{\kappa}_k^0$ and $\boldsymbol{\mu}_k^0$ are obtained according to Eq. (19). One can also obtain the last term of Eq. (A2) as

$$\frac{\partial J}{\partial \varphi_i} = \frac{1}{w_{elec}}\left(\frac{1}{2}\int_D \boldsymbol{\varepsilon}^T \frac{\partial \boldsymbol{C}}{\partial \varphi_i} \boldsymbol{\varepsilon}\, d\Omega\right) - \frac{w_{\text{mech}}}{w_{elec}^2}\left(\frac{1}{2}\int_D \boldsymbol{E}^T \frac{\partial \boldsymbol{\kappa}}{\partial \varphi_i} \boldsymbol{E}\, d\Omega\right) \quad \text{(A11)}$$

For the case of four material phases, $\frac{dV_k}{d\varphi_i}$ for $i = 1,2$ and $k = 1,2,3,4$ is obtained by

$$\frac{-dV_3}{d\varphi_1} = \frac{-\partial V_3}{\partial \varphi_1} = \frac{dV_1}{d\varphi_1} = \frac{\partial V_1}{\partial \varphi_1} = \int_D \frac{\partial \rho_1}{\partial \varphi_1}\, d\Omega \approx \sum_e \widetilde{H}(\Phi_2(X_e))\widetilde{\delta}(\Phi_1(X_e))\frac{\partial \Phi_1(X_e)}{\partial \varphi_1} \quad \text{(A12.a)}$$

$$\frac{-dV_4}{d\varphi_1} = \frac{-\partial V_4}{\partial \varphi_1} = \frac{dV_2}{d\varphi_1} = \frac{\partial V_2}{\partial \varphi_1} = \int_D \frac{\partial \rho_2}{\partial \varphi_1}\, d\Omega \approx \sum_e \left(1 - \widetilde{H}(\Phi_2(X_e))\right)\widetilde{\delta}(\Phi_1(X_e))\frac{\partial \Phi_1(X_e)}{\partial \varphi_1} \quad \text{(A12.b)}$$

$$\frac{-dV_2}{d\varphi_2} = \frac{-\partial V_2}{\partial \varphi_2} = \frac{dV_1}{d\varphi_2} = \frac{\partial V_1}{\partial \varphi_2} = \int_D \frac{\partial \rho_1}{\partial \varphi_2}\, d\Omega \approx \sum_e \widetilde{H}(\Phi_1(X_e))\widetilde{\delta}(\Phi_2(X_e))\frac{\partial \Phi_2(X_e)}{\partial \varphi_2} \quad \text{(A12.c)}$$

$$\frac{-dV_4}{d\varphi_4} = \frac{-\partial V_4}{\partial \varphi_4} = \frac{dV_3}{d\varphi_2} = \frac{\partial V_3}{\partial \varphi_2} = \int_D \frac{\partial \rho_3}{\partial \varphi_2} d\Omega \approx \sum_e \left(1 - \widetilde{H}(\Phi_1(X_e))\right) \tilde{\delta}(\Phi_2(X_e)) \frac{\partial \Phi_2(X_e)}{\partial \varphi_2} \quad \text{(A12.d)}$$

and for the case of three material phases, one can write

$$\frac{dV_{solid}}{d\varphi_1} = \frac{\partial V_{solid}}{\partial \varphi_1} = \int_D \frac{\partial \rho_1}{\partial \varphi_1} d\Omega \approx \sum_e \tilde{\delta}(\Phi_1(X_e)) \frac{\partial \Phi_1(X_e)}{\partial \varphi_1} \quad \text{(A13.a)}$$

$$\frac{dV_{flexo}}{d\varphi_1} = \frac{\partial V_{\text{flexo}}}{\partial \varphi_1} = \int_D \frac{\partial \rho_2}{\partial \varphi_1} d\Omega \approx \sum_e \widetilde{H}(\Phi_2(X_e)) \tilde{\delta}(\Phi_1(X_e)) \frac{\partial \Phi_1(X_e)}{\partial \varphi_1} \quad \text{(A13.b)}$$

$$\frac{dV_{flexo}}{d\varphi_2} = \frac{\partial V_{\text{flexo}}}{\partial \varphi_2} = \int_D \frac{\partial \rho_2}{\partial \varphi_2} d\Omega \approx \widetilde{H}(\Phi_1(X_e)) \tilde{\delta}(\Phi_2(X_e)) \frac{\partial \Phi_2(X_e)}{\partial \varphi_2} \quad \text{(A13.c)}$$

where $\tilde{\delta}(\Phi_i) = \frac{\partial \widetilde{H}(\Phi_i)}{\partial \Phi_i}$ is the approximate Dirac delta function defined as

$$\tilde{\delta}(\Phi_i) = \begin{cases} \frac{3}{4\Delta}\left(1 - \left(\frac{\Phi_i}{\Delta}\right)^2\right) & \text{for } -\Delta \leq \Phi_i \leq \Delta \\ 0 & \text{otherwise} \end{cases} \quad \text{(A14)}$$

and $\frac{\partial \Phi_i(X_e)}{\partial \varphi_i}$ is calculated by

$$\frac{\partial \Phi_i(X_e)}{\partial \varphi_i} = N_{i,j}^{p,q}(\xi, \eta) \quad \text{(A15)}$$

**References:**

1- S. Nanthakumar, T. Lahmer, X. Zhuang, G. Zi, T. Rabczuk, Detection of material interfaces using a regularized level set method in piezoelectric structures, Inverse Problems in Science and Engineering 24(1) (2016) 153–176.

2- S. S. Nanthakumar, T. Lahmer, T. Rabczuk, Detection of multiple flaws in piezoelectric structures using XFEM and Level sets, Computer Methods in Applied Mechanics and Engineering 275 (2014) 98-112.

3- S. S. Nanthakumar, T. Lahmer, T. Rabczuk, Detection of flaws in piezoelectric structures using XFEM, International Journal of Numerical Methods in Engineering 96(6) (2013) 373–389.

4- N.D. Sharma, R. Maranganti, P. Sharma, On the possibility of piezoelectric nanocomposites without using piezoelectric materials, Journal of the Mechanics and Physics of Solids 55 (2007) 2328–2350.

5- P. V. Yudin, A. K. Tagantsev, TOPICAL REVIEW: Fundamentals of flexoelectricity in solids, Nanotechnology 24 (2013) 432001 (36pp), doi:10.1088/0957-4484/24/43/432001.

6- H. Cao, V. Leung, C. Chow, H. Chan, Enabling technologies for wireless body area networks: A survey and outlook, IEEE Communications Magazine 47(12) (2009) 84-93.

7- D. Sanders, Environmental sensors and networks of sensors, Sensor Review 28 (2008) 273-274.

8- 8- S. Kim, S. Pakzad, D. Culler, J. Demmel, G. Fenves, S. Glaser, M. Turon, Health monitoring of civil infrastructures using wireless sensor networks, Proceedings of the Sixth International Symposium on Information Processing in Sensor Networks (2007) 254 –263.

9- Z. L. Wang, W. Wu, Nanotechnology-enabled energy harvesting for self-powered micro-/nanosystems, Angewandte Chemie International Edition 51 (2012) 11641–11903.

10- H. Ghasemi, H. S. Park, T. Rabczuk, A level-set based IGA formulation for topology optimization of flexoelectric materials, Comput. Methods Appl. Mech. Engrg 313 (2017) 239-258.

11- S.S. Nanthakumar, X. Zhuang, H.S. Park, T. Rabczuk, Topology optimization of flexoelectric structures, Journal of the Mechanics and Physics of Solids, 105 (2017) 217-234.

12- E. Silva, S. Nishiwaki, N. Kikuchi, Topology optimization design of flextensional actuators, IEEE Transactions on Ultrasonics, Ferroelectrics and Frequency Control 47 (2000) 657–671.

13- Q. Nguyen, L. Tong, Voltage and evolutionary piezoelectric actuator design optimization for static shape control of smart plate structures, Materials and Design 28 (2007) 387–399.

14- A. Takezawa, M. Kitamura, S. L. Vatanabe, Design methodology of piezoelectric energy-harvesting skin using topology optimization, Struct Multidisc Optim. 49 (2014) 281-297.

15- GIN Rozvany, M. Zhou, T. Birker, Generalized shape optimization without homogenization, Struct Multidisc Optim 4 (1992) 250–254.

16- Amir Abdollahi, Christian Peco, Daniel Millan, Marino Arroyo, Irene Arias, Computational evaluation of the flexoelectric effect in dielectric solids, JOURNAL OF APPLIED PHYSICS 116 (2014) 093502.

17- M. S. Majdoub, P. Sharma, T. Cagin, Enhanced size-dependent piezoelectricity and elasticity in nanostructures due to the flexoelectric effect, PHYSICAL REVIEW B 79 (2009) 119904 (E).

18- S. Shen, S. Hu, A theory of flexoelectricity with surface effect for elastic dielectrics, J. of the mechanics and physics of solids 58 (2010) 665-677.

19- Michael Yu Wang, Xiaoming Wang, Dongming Guo, A level set method for structural topology optimization, Comput. Methods Appl. Mech. Engrg. 192 (2003) 227-246.

20- H. K. Zhao, B. Merriman, S. Osher, L. Wang, Capturing the behavior of bubbles and drops using the variational level set approach, J. Comput. Phys. 143 (1998) 495–518.

21- M. Wang, X. Wang, ''Color'' level sets: a multi-phase method for structural topology optimization with multiple materials, Comput. Methods Appl. Mech. Engrg. 193 (2004) 469–496.

22- L.A. Vese, T.F. Chan, A multiphase level set framework for image segmentation using the Mumford and Shah model, Int. J. Comput. Vision 50 (3) (2002) 271–293.

23- J. Luo, Z. Luo, L. Chen, L. Tong, MY. Wang, A semi implicit level set method for structural shape and topology optimization, J Comput Phys 227 (2008) 5561-5581.